\documentclass{appolb}
\usepackage{epsfig,dcolumn}
\usepackage{graphicx}
\usepackage{color}

\usepackage{amsmath}
\usepackage{amsfonts}
\usepackage{amssymb}
\usepackage{graphicx}
\usepackage{type1cm}
\usepackage{eso-pic}

\newcommand{\be}{\begin{equation}}
\newcommand{\ee}{\end{equation}}

\usepackage{bm} 

\begin{document}

\title{Calculation of Regge trajectories of strange resonances and identification of the $K_0^*(800)$ as a non-ordinary meson%
\thanks{Presented at Excited QCD 2016, Costa do Caparica, Portugal, March 6-12}%
}
\author{J.R.Pel\'aez, A.Rodas 
\address{Departamento de F\'isica Te\'orica II, Universidad Complutense de Madrid, 28040 Madrid, Spain}
\\
}
\maketitle
\begin{abstract}
We review how the Regge trajectory of an elastic resonance can be
obtained just from its pole position and coupling, using a dispersive
formalism. This allows us to deal correctly with the finite widths of resonances
in Regge trajectories. In this way we can calculate the Regge trajectories for the $K^*(892)$, $K_1(1400)$ and $K^*_0(1430)$, obtaining ordinary linear Regge trajectories, expected for $q \bar q$ resonances. In contrast, for
the $K^*_0(800)$ meson, the resulting Regge trajectory is non-linear and with
much smaller slope, strongly supporting its non-ordinary nature.
\end{abstract}

\section{Introduction}

In recent works \cite{Londergan:2013dza, Carrasco:2015fva} we have used the analytic properties of amplitudes in the complex angular momentum plane to study the Regge trajectories of resonances decaying predominantly to one channel. One of the main features of this calculation is the correct relation of the resonance width to the imaginary part of the trajectory. In principle, the form of these trajectories can be used to discriminate between the underlying QCD mechanisms that generate
these resonances. Ordinary linear $(J,M^2)$ trajectories relating the angular
momentum J and the mass squared are intuitively interpreted in terms of
$q \bar q$ states, since they can be easily obtained using  a rotating flux tube model. Strong deviations from
this linear behavior would suggest a rather different nature and the scale of
the trajectory would also indicate the scale of the mechanism responsible for
the formation of the resonance.
 
In particular, we have studied in \cite{Londergan:2013dza} and \cite{Carrasco:2015fva} several resonances appearing in $\pi \pi$  or $K \bar K$scattering, including the  $\rho(770)$, $f_2(1275)$, $f'_2(1525)$, whose resulting Regge trajectory are ordinary linear $(J,M^2)$ trajectory, whereas the $f_0(500)$ Regge trajectory does not follow the ordinary pattern \cite{Anisovich:2000kxa}. Actually, at very low energy its trajectory resembled that of the familiar Yukawa potentials \cite{yukawa}. Here we include our preliminary results on strange elastic resonances, which can be studied almost in the same way. We will see that the resulting trajectories for the $K^*(892)$,$K_1(1400)$ and $K^*_0(1430)$ are linear and with ordinary slopes, whereas the $K^*_0(800)$ has almost the same behavior as the $f_0(500)$ in the low energy region $M^2<2$ GeV$^2$. This might be considered an indication of its non-ordinary nature.

\section{Regge trajectories}

Near a resonance pole the partial wave reads

\be
t_l(s)  = \frac{\,\beta(s)}{l-\alpha(s)\,} + f(l,s),
\label{reggepw}
\ee
where $f(l,s)$ is an analytical function near $l=\alpha(s)$. If we consider now that the pole dominates the partial wave behavior the unitarity condition $\mbox{Im}t_l(s)=\rho(s)|t_l(s)|^2$ implies that

\be
\mbox{Im}\,\alpha(s)   = \rho(s) \beta(s).   
\label{uni} 
\ee

Furthermore, the Regge trajectory $\alpha(s)$ and residue $\beta(s)$ satisfy the Schwarz reflection principle, {\it i.e.}, $\alpha(s^*)=\alpha^*(s)$ and $\beta(s^*)=\beta^*(s)$. The analytic properties of $\alpha(s)$ and $\beta(s)$ and the elastic unitarity condition imply the following system of coupled dispersion relations \cite{Chu:1969ga, Reggeintro}

\begin{align}
\mbox{Re} \,\alpha(s) & =   \alpha_0 + \alpha' s +  \frac{s}{\pi} PV \int_{4m^2}^\infty ds' \frac{ \mbox{Im}\alpha(s')}{s' (s' -s)}, \label{iteration1}\\
\mbox{Im}\,\alpha(s)&=  \frac{ \rho(s)  b_0 \hat s^{\alpha_0 + \alpha' s} }{|\Gamma(\alpha(s) + \frac{3}{2})|}
 \exp\Bigg( - \alpha' s[1-\log(\alpha' s_0)] \nonumber \\
&+  \!\frac{s}{\pi} PV\!\!\!\int_{4m^2}^\infty\!\!\!\!\!\!\!ds' \frac{ \mbox{Im}\alpha(s') \log\frac{\hat s}{\hat s'} + \mbox{arg }\Gamma\left(\alpha(s')+\frac{3}{2}\right)}{s' (s' - s)} \Bigg), 
\label{iteration2}\\
 \beta(s) &=    \frac{ b_0\hat s^{\alpha_0 + \alpha' s}}{\Gamma(\alpha(s) + \frac{3}{2})} 
 \exp\Bigg( -\alpha' s[1-\log(\alpha' s_0)] \nonumber \\
&+  \frac{s}{\pi} \int_{4m^2}^\infty \!\!\!\!\!\!\!ds' \frac{  \mbox{Im}\alpha(s') \log\frac{\hat s}{\hat s'}  + \mbox{arg }\Gamma\left(\alpha(s')+\frac{3}{2}\right)}{s' (s' - s)} \Bigg),
 \label{betafromalpha}
 \end{align}
where $PV$ denotes the principal value.  For real $s$, the last two equations reduce to Eq.\eqref{uni}.
The three equations are solved numerically with the free parameters fixed by demanding 
 that the pole on the second sheet of the amplitude in Eq.~(\ref{reggepw}) is at the pole positions associated to the resonance under study.

For the $K^*_0(800)$, one should also
make explicit in $\beta(s)$ the Adler-zero required by the chiral symmetry. In
that case, $b_0$ will not be dimensionless.

\begin{figure}
\centering
\centerline{\includegraphics[width=0.48\linewidth]{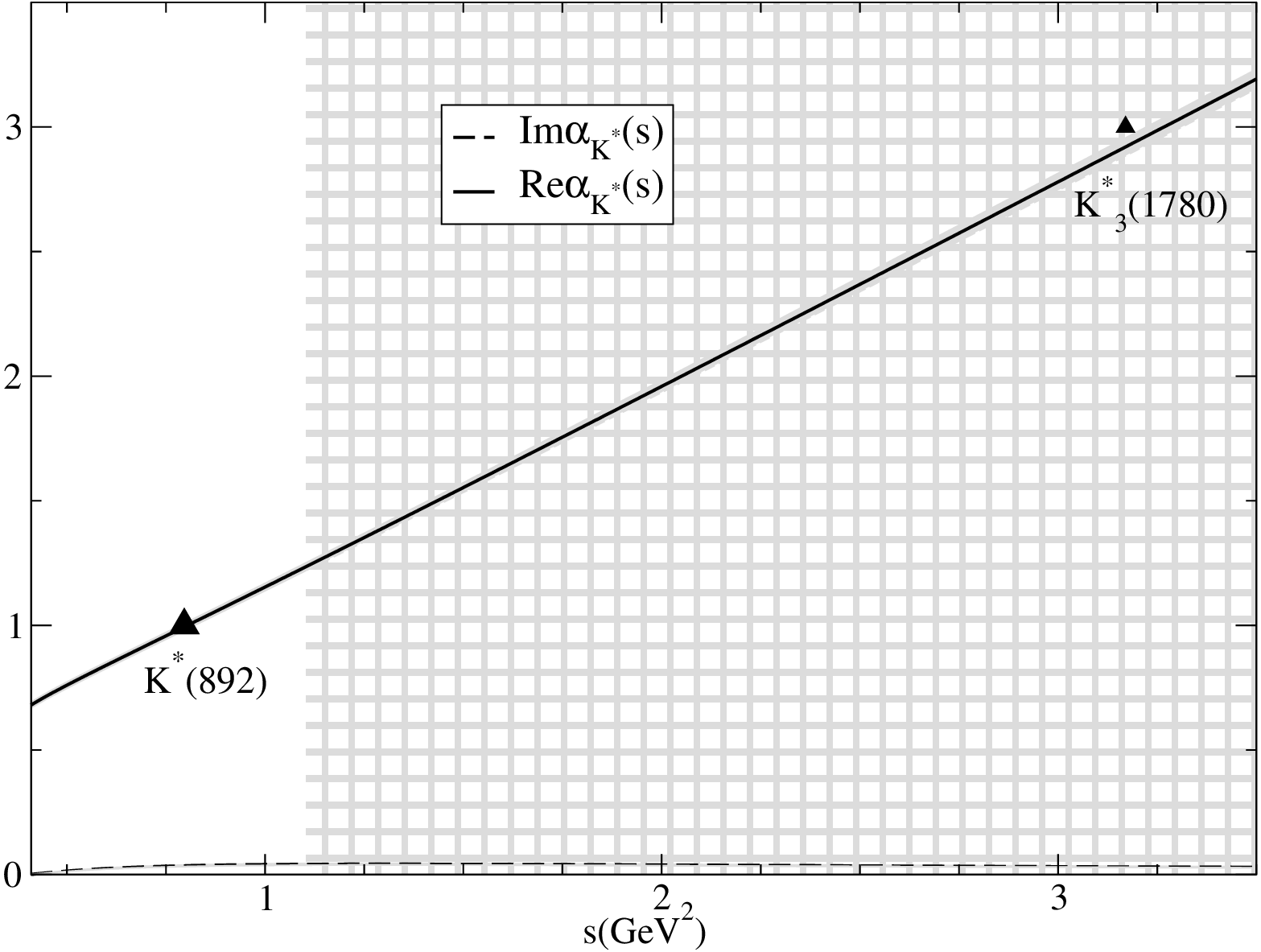}   \includegraphics[width=0.48\linewidth]{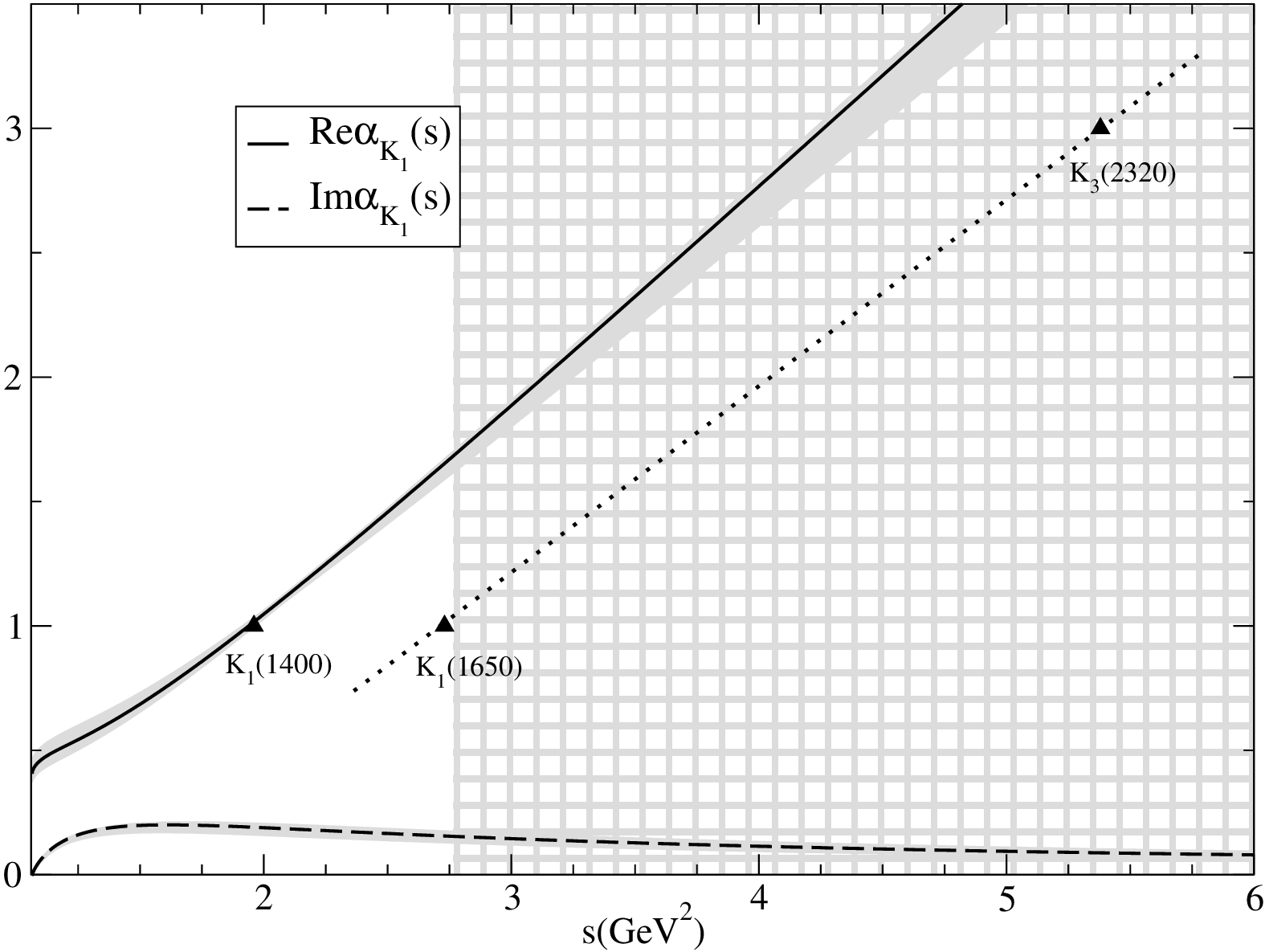}}
\vspace*{.2cm}
\centerline{\includegraphics[width=0.48\linewidth]{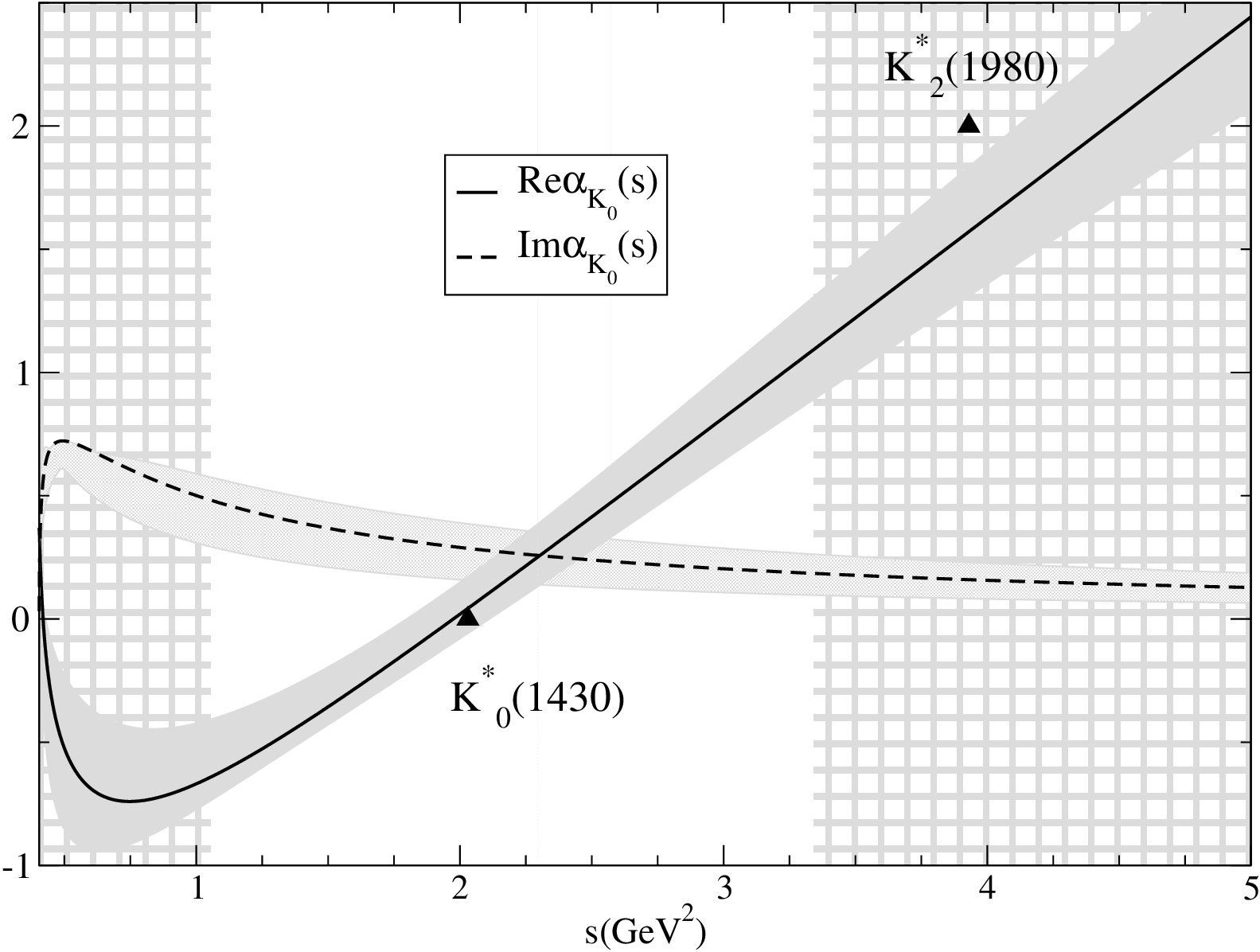}  \includegraphics[width=0.48\linewidth]{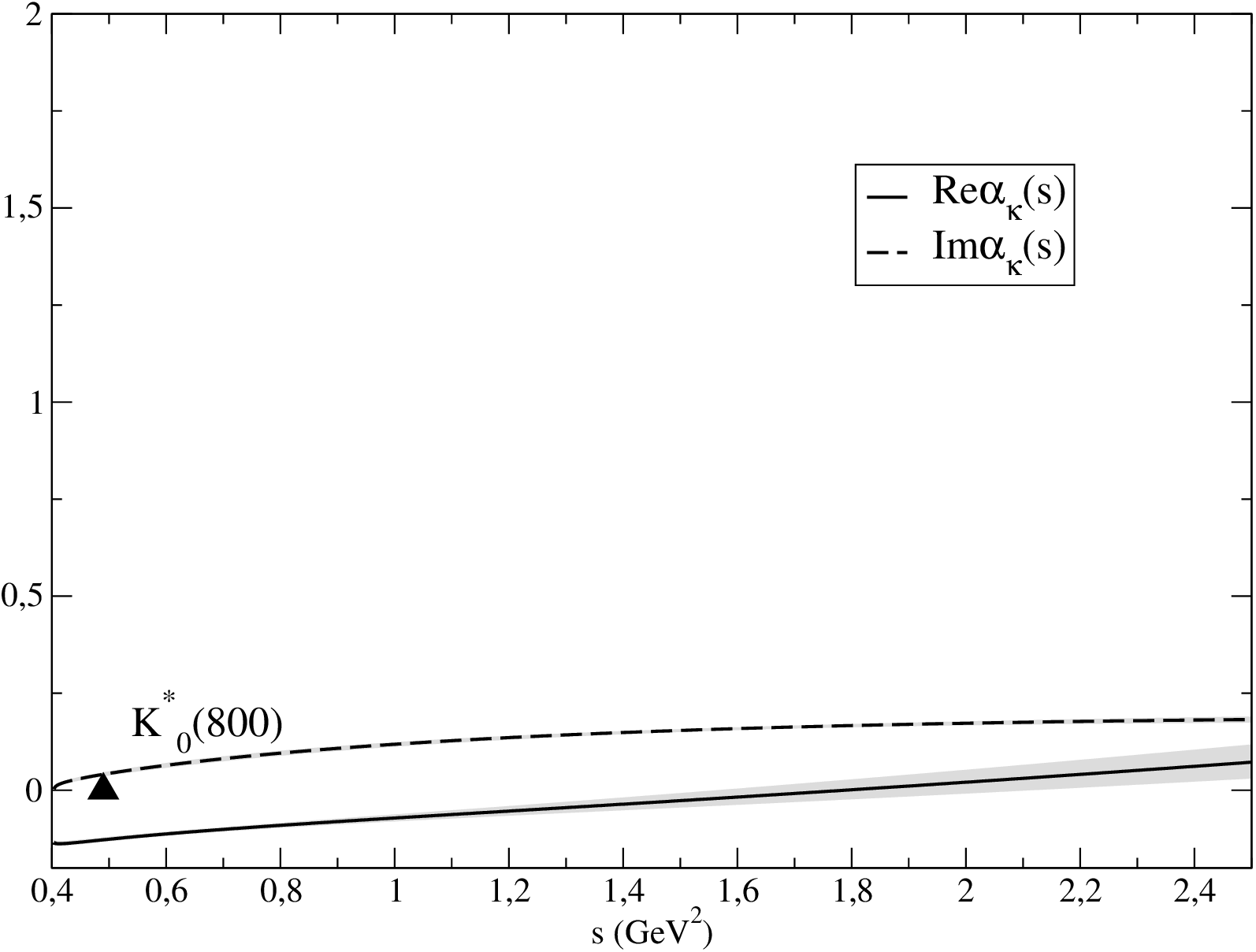}}
\caption{\rm \label{fig:Regge} Regge trajectories calculated from their
poles. The continuous lines correspond to the
real part of the trajectory (to be identified with spin at integer values), whereas the
dashed lines stand for the imaginary parts. The gray bands cover the uncertainties
in our calculation, mostly due to using the elastic approximation. The light gray area is the mass region where
our approach should be considered cautiously as a mere extrapolation.
}
\end{figure}

We solve the system of Eqs.~(\ref{iteration2}) and (\ref{betafromalpha}) iteratively. The values of the parameters
are fixed by fitting only three inputs, namely, the real and imaginary parts
of the resonance pole position $s_M=(M_R-i\Gamma_R/2)^2$, where $M_R$ and $\Gamma_R$ are
the pole mass and width of the resonance, together with the absolute value
of the pole residue $|g_M|$. Namely, we fit the resonance pole on the second
Riemann sheet to: $\beta_M(s)/(l-\alpha_M(s)) \rightarrow |g_M^2|/(s-s_M)$.

The pole parameters of the $K^*(892)$ and the $K^*_0(800)$ are taken from a dispersive analysis of $\pi K$ scattering \cite{Pelaez:2016tgi}. For the $K_1(1400)$ and the $K^*_0(1430)$ we use a usual Breit-Wigner description, taking their values from the Review of Particle Physics \cite{PDG}.

\begin{table} 
\caption{Parameters of the $K^*(892)$, $K_1(1400)$, $K^*_0(1430)$ and $K^*_0(800)$ Regge trajectories calculated from their poles
in meson-meson scattering. For the $K^*_0(1430)$ and $K^*_0(800)$, $b_0$ is not dimensionless because we have factorized
explicitly in $\beta(s)$ the Adler zero required by chiral symmetry. Note the similar $\alpha'$ for all resonances except $K^*_0(800)$} 
\centering 
\begin{tabular}{c | c c c} 
\hline\hline  
\rule[-0.15cm]{0cm}{.55cm}& $\alpha_0$ & $\alpha'$ [GeV$^-2$] & $b_0$\\
\hline 
\rule[-0.15cm]{0cm}{.55cm} $K^*(892)$&0.32$\pm$0.01  & 0.83$\pm$0.01 & 0.48$\pm$0.03\\
\rule[-0.15cm]{0cm}{.55cm} $K_1(1400)$&$-0.72^{+0.13}_{-0.03}$  & $0.90^{+0.01}_{-0.07}$ & $6.02^{+0.39}_{-1.13}$\\
\rule[-0.15cm]{0cm}{.55cm} $K^*_0(1430)$&$-1.15^{+0.23}_{-0.15}$& $0.81^{+0.08}_{-0.1}$ & $4.04^{+1.26}_{-2.43}$\\
\rule[-0.15cm]{0cm}{.55cm} $K^*_0(800)$&0.28$\pm$0.02  & 0.15$\pm$0.01 & 0.44$\pm$0.04\\
\hline 
\end{tabular} 
\label{tab:parameters} 
\end{table}

In Fig.\ref{fig:Regge} we show the resulting trajectories. The resulting parameters are given in Table~\ref{tab:parameters}. For the $K^*(892)$,$K_1(1400)$ and $K^*_0(1430)$ trajectories the imaginary part is much smaller than the real part near the resonances, growing linearly in s with a typical slope for Regge trajectories of $\alpha'\simeq 0.9$ GeV$^{-2}$.

In contrast, as one can observe in Table \ref{tab:parameters}, the $K^*_0(800)$ slope of the resulting curve is almost one order of magnitude smaller than that of ordinary mesons. This provides support for the non-ordinary nature of the resonance. The tiny trajectory excludes that any of the known isoscalar resonances may lie on the $K^*_0(800)$ trajectory. We have also tried to force a typical linear trajectory for the $K^*_0(800)$, but that completely spoils the data description and the $K^*_0(800)$ pole fit. Hence the approximation doesn't hold near the resonance and the method is not applicable.

\begin{figure}
\centering
\includegraphics[scale=0.4]{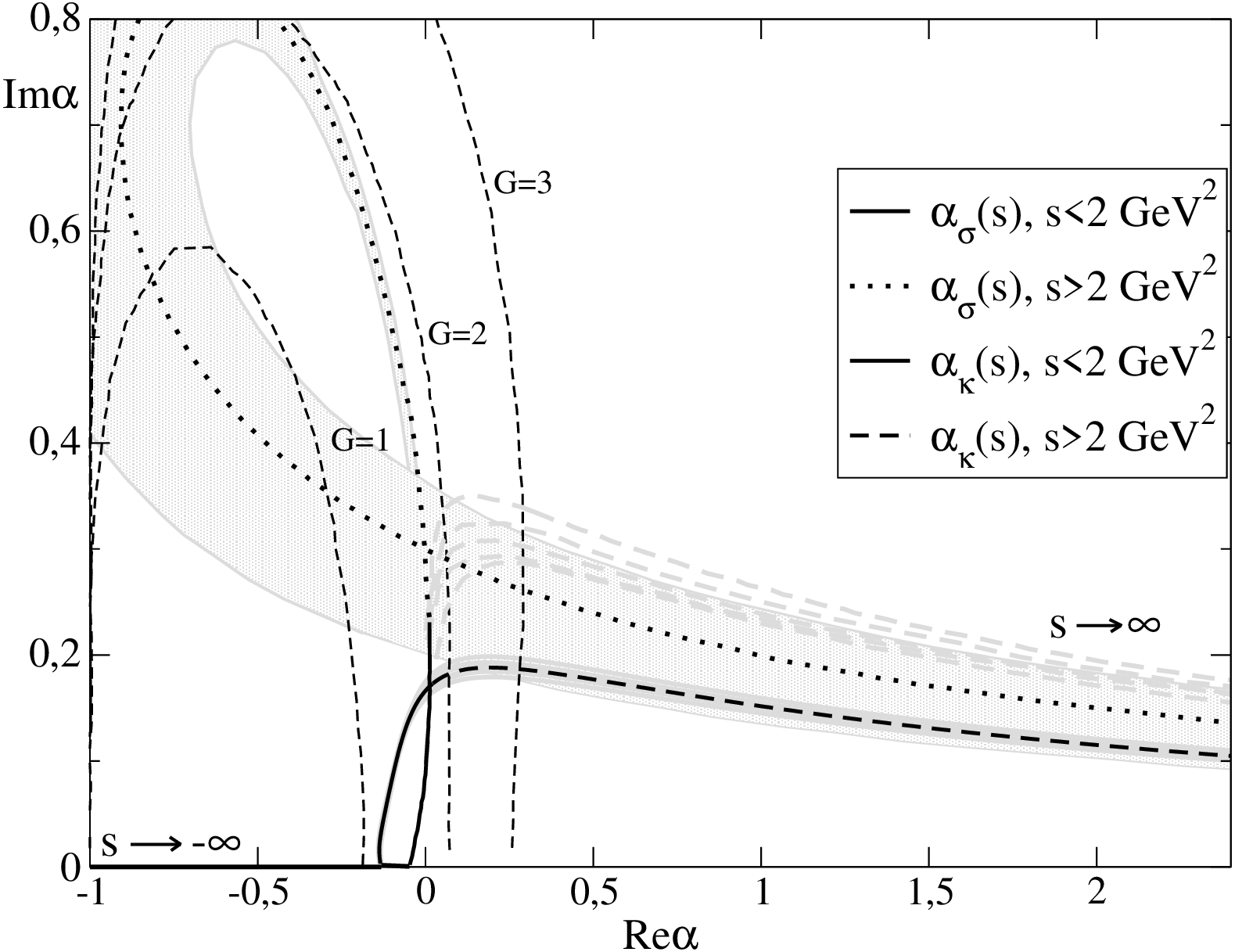} 
\caption{\rm \label{fig:comparison} 
At low and intermediate energies (thick continuous lines), the trajectories of the $f_0(500)$ and the $K^*_0(800)$ are similar
to those of Yukawa potentials $V(r)= Ga \exp(-r/a)/r$ \cite{yukawa}  (thin dashed lines).
Beyond 2 GeV$^2$, we plot our results as thick discontinuous lines because they should
be considered just as extrapolations.
}
\end{figure}

In Fig.\ref{fig:comparison} we show the similarities between the $K^*_0(800)$ and the $f_0(500)$ trajectories, both also very similar to those of Yukawa potentials in non-relativistic scattering \cite{yukawa}. The trajectory of the $f_0(500)$ is almost equal to the $G=2$ Yukawa curve up to $s=2$ GeV$^2$, while the curve with $G=1.5$ is near the curve of the $\kappa$. Hence we can estimate $a_{\pi \pi}=0.5$ GeV$^{-1}$ and $a_{\pi K}=0.3$ GeV$^{-1}$ being $a_{\pi \pi}/a_{\pi K} \approx \mu_{\pi K}/\mu_{\pi \pi}$. Being the resonance a kind of a molecular state between two mesons with a short-range potential.

In summary, our formalism is able to predict the Regge trajectory of a resonance using just the parameters associated to the pole while describing the amplitude near the resonance. We have calculated the Regge trajectories of the $K^*(892)$, $K_1(1400)$ and $K^*_0(1430)$ obtaining similar results as in \cite{Londergan:2013dza, Carrasco:2015fva} for the $\rho(770)$, $f_2(1270)$ and $f'_2(1525)$ typical $q \bar q$ behavior. However this is not the case for the $K^*_0(800)$, which comes out to be similar to the $f_0(500)$ meson.

\section{Acknowledgements}
J.R.P. and A.R. are supported by Spanish Projects No. FPA2011-27853-C02-02 and No. FPA2014-53375C2-2 and Red de Excelencia de Física Hadrónica FIS2014-57026-REDT.

\end{document}